\documentstyle[preprint,floats,prl,aps,epsf]{revtex}

\begin{document}

\preprint{\vbox{\hbox {April 1999} \hbox{IFP-772-UNC} }}

\draft
\title{Mersenne Primes, Polygonal Anomalies and String Theories Classification}
\author{\bf Paul H. Frampton$^{(a)}$ and Thomas W. Kephart$^{(b)}$}
\address{(a)Department of Physics and Astronomy,\\
University of North Carolina, Chapel Hill, NC  27599.}
\address{(b)Department of Physics and Astronomy,\\
 Vanderbilt University, Nashville, TN 37325.}
\maketitle
\date{\today}

\begin{abstract}
It is pointed out that the Mersenne primes $M_p=(2^p-1)$ and associated 
perfect numbers ${\cal M}_p=2^{p-1}M_p$ play a significant 
role in string theory; this observation may suggest
a classification of consistent string theories.
\end{abstract}
\pacs{}

\newpage

Anomalies and their avoidance have provided a guidepost in constraining
viable particle physics theories. From the standard model to
superstrings,
the importance of finding models where the concelation of local and
global
anomalies that spoil local invariance properties of theories, and hence
render them inconsistent, cannot be overestimated. The fact that
anomalous
thories can be dropped from contention has made progress toward the true
theory of elementary particles proceed at an enormously accelerated
rate.
Here we take up a systematic search, informed by previous results and as
yet
partially understood connections to number theory, for theories free of
leading gauge anomalies in higher dimensions. We will find new cases and
be
able to place previous results in perspective.

In number theory a very important role is played by the Mersenne primes
$%
M_{p}$ based on the formula

\begin{equation}
M_p = 2^p - 1  \label{mersenne}
\end{equation}

\noindent where $p$ is a prime number. $M_p$ is sometimes itself a prime
number. The first 33 such Mersenne primes correspond\cite{HW,ito,web} to
prime numbers below one million:
\begin{eqnarray}
p & = & 2, 3, 5, 7, 13, 17, 19, 31, 61, 89, 107, 127, 521, 607, 1279,
2203,
2281,  \nonumber \\
& & 3217, 4253, 4423, 9689, 9941, 11213, 19937, 21701, 23209, 44497,
\nonumber \\
& & 86243, 110503, 132049, 216091, 756839, 859433.  \label{primes}
\end{eqnarray}
As a comparison to this remarkable sequence of the first 33 Mersenne
primes,
there are altogether 78498 primes below one million so that Eq.(\ref
{mersenne}), although an invaluable source of large prime numbers, far
more
often generates a composite number than a prime.

On the occasion that Eq.(\ref{mersenne}) {\it does} generate a prime, an
immediate derivative thereof is the perfect number which we shall
designate $%
{\cal M}_p$ given by ${\cal M}_p = 2^{p-1}M_p$. It is straightforward
and
pleasurable to prove in general that ${\cal M}_p$ is {\it perfect},
defined
as ${\cal M}_p$ equalling the sum of all of its divisors. For example,
$%
{\cal M}_2=6=1+2+3$, ${\cal M}_3=28=1+2+4+7+14$, and so on. The ${\cal
M}_p$
are the only even perfect numbers; it is unknown if there is an odd
perfect
number but if there is one it is known \cite{BCt} that it is larger than
$%
10^{300}$.

In the present Letter, we shall associate the perfect numbers derived
from
Mersenne primes with the polygonal anomalies whose cancellation
underlies
the successful string theories.

For example, heterotic and type-I superstrings in ten dimensions are
selected to have gauge groups $O(32)$ and $E(8) \times E(8)$ on the
basis of
anomaly cancellation of the hexagon anomaly\cite{FK1,FK2,FK3,TS}.
Equivalently, these two superstrings correspond to the only self-dual
lattices in 16 dimensions: $\Gamma_8 \bigoplus \Gamma_8$ and
$\Gamma_{16}$%
\cite{serre}. The dimension of these two acceptable gauge groups in
$d=10$
is $dim(G)=496={\cal M}_5$, indeed a perfect number of the Mersenne
sequence. Further motivation in low dimensions for consideration of the
perfect number comes from \cite{HolmanKephart} ${\cal M}_3$ the $SO(8)$
and $%
G_{2} \times G_{2}$ supergravities in 6 dimensions for ${\cal M}_3$ ,
from
noting that $0(4)$ and $SU(2)\times SU(2)$ are anomaly free in four
dimensions for ${\cal M}_2$ and  from the existence of an ${\cal N}=2$
world
sheet supersymmetric string theory in $2$ dimensions \cite{2D}with gauge
group
$SO(2)\sim U(1)$ for ${\cal M}_1$.

The appropriate polygon for spacetime dimension $d$ is the $l-$agon
where $l
= (\frac{d}{2}+1)$.

One way to discover the significance of $M_p$ and ${\cal M}_p$ in string
theory is to recognize that the leading $l-$agon anomaly for a k-rank
tensor
of $SU(N)$ or $O(N)$ is given \cite{FK1,FK3} by a generalized Eulerian
number ( the Eulerian numbers are $A_N(N,k)$)
\begin{equation}
A_l(N,k) = \sum_{p=1}^{k-1}(-1)^{k-p-1}(k-p)^{l-1}\left(
\begin{array}{c}
N \\
p
\end{array}
\right)  \label{eulerian}
\end{equation}

Our purpose here is to investigate the space-time dimensions
corresponding
to the Mersenne primes $D=2p$ for gauge group irreps with vanishing
leading
gauge anomalies. One could then cancel the nonleading anomalies in the
Green-Schwarz mechanism to generate a candidate string theory or
supergravity (a complete theory must also avoid all local gravitational
and
global anomalies).

Since all the primes except $2$ are odd, the Mersenne prime dimensions
(MPDs) are $D=4n+2$, where $n$ is an integer except for the special case
$D=4$. A thorough investigation of the MPDs returns the following for
$D$
between 4 and 26 and certain higher values:

\bigskip

{\bf $D=4$} is well studied in the literature, to which we refer the
reader
\cite{Slansky}.

\bigskip

{\bf $D=6$} ($p=3$): Expressing antisymmetric tensor irreps by $[1]^k$
we
again find anomaly freedom for the second rank antisymmetric tensor
$k=2$
when $N=8$ for gauge groups $SU(8)$ or $SO(8)$. For $SU(N)$ one expects
the
conjugate solution $[1]^{N-k}=[1]^{8-2}=[1]^{6}$, which is nothing new,
but
due to its low order, the anomaly polynomial factorizes at $[1]^{6}$ to
$%
(n-6)(N-27)$, implying a nonvanishing anomaly of $[1]^{6}$ for $SU(27)$
[and
for $SO(27)$]. This in turn implies a $[1]^{N-k}=[1]^{27-6}=[1]^{21}$
solution, which one finds at $N=98$. This sequence continues, (see Table
1) (We remind the reader that for $SU(N),$ $[1]^{N-k}$ and $[1]^{k}$ are
complex irreps except when $k=N/2$ and $N$ is even where $[1]^{N/2}$ is
a
real irrep. For $SO(N)$, $[1]^{k}$ is real for $k<N/2$. When $k=N/2$, 
$[1]^{k}$ splits. The components are real if $N/2$ is even and they are a
complex conjuate pair if $N/2$ is odd.There are added subtlities for
$SO(8)$
because of triality \cite{Slansky}.)

\bigskip

$D=10$. As with $D=6$ we find a $[1]^{2}$ solution when $N=2p=32$. There
are
two further solutions, up to conjugation $k=4$ with $N=12$, and $k=10$
with $%
N=32$, and no others with $k\leq 40$. (In what follows we do a study of
all
cases out to $k=40$, unless noted otherwise.)

\bigskip

$D=14$: The only solution is $k=2$ for $N=128$.

\bigskip

The case where $p=13$ deserves special consideration, since it
corresponds
to 26 dimensions, and a $26D$ theory with $SO(2^{13})=SO(8192)$ has
indeed
already been considered in the literature \cite{DG,W,BS}. In \cite{DG},
the
single dilaton emission amplitude from a disk world sheet was calculated
and
used in a proof that the total dilaton emission amplitude (from the
projective plane plus the disk \cite{GW}) at this order vanishes in
$26D$
for $SO(8192)$. Furthermore, it has been shown \cite{W} that the
one-loop
divergences are avoided by $SO(8192)$ open strings in $26D$. A general
understanding has been provided \cite{BS} of the Chan-Paton factors
for$SO(2^{D/2})$ in terms of D added fermionic variables at the ends of open
strings, and this is useful input into developing the partition function
for
the $SO(8192)$ open string \cite{BS}.

Likewise, the only solution is $k=2$ for $N=2^{p}$, with $%
D=2p=34,38,62,178,214,254,1042,1214,2558$ and $4406$, where we have
searched
through $k=40$ except for $D=1042$, where $k\leq 24$, $D=2558$ where
$k\leq %
10$, and $D=4406$ where $k\leq 8.$

\bigskip

For the sake of completeness, we have also studied the remaining even
dimensions below $D=26$, with no Mersenne prime correspondence. As
before, $%
k=2$ with $N=2^{p}$ is always a solution, and when $D=4n$ ($n$ integer),

$k=N/2$ is also a solution as expected since it is real. (Recall that
real
representations have no anomalies in $D=4n$ dimensions, but do in
$D=4n+2$,
therefore anomaly freedom for [1]$^{k}$ irreps is trivial in $D=4n$ for
$SO$
groups, but not for $SU$ groups.)

$D=8$: We find the usual $k=2$ and $k=N/2$ solutions, plus two more
sequences, one starting with $k=2$, $D=16,$ and the other with $k=3$,
$D=27$%
(see Table 1).

$D=12$: Has only $k=2$ and $k=N/2$ solutions.

$D=16$: Has the usual $k=2$ and $k=N/2$ solutions, plus at $k=3$ with
$N=27$
and also at $k=3$ with $N=486$.

$D=18,20,22$ and $24$ have nothing new beyond the usual solutions, of
$k=2$
and $k=N-2$ for $SU(N^p)$ with $p=D/2$, and for $D=4n$ the trivial case
of
the real representation $k=N/2$ for any $SU(N)$.

\bigskip

This completes the classification.

\bigskip

Returning to $D=8,$ the $[1]^{3}=2925$ of $SU(27)$ or $SO(27)$ is
anomaly
free, but also the $[1]^{3}$ of $E_{6}$ is a $2925$ under the
decomposition $%
SU(27)\longrightarrow E_{6}$, where $27\longrightarrow 27$. Since the
generalized Casimir invariants of $E_{6}$ are of rank $2,5,6,8,9$, and
$12$,
leading anomalies are expected at $D=2,8,20,14,16$, and $22$.
\cite{KVaughn}%
. However, the $2925$ is an exception since it is real.

\bigskip

In $D=6$ no leading $E_{6}$ anomalies are expected, and we find the
$k=6$, $%
N=27$ result corresponding to the $[1]^{6}=296010$ irrep of $SU(27)$ or
$%
SO(27)$ is reducible in $E_{6}$.

\bigskip

In $D=16$ for $k=3$ and $N=27$, leading $E_{6}$ anomalies are voided by
the $%
2925$.

The higher $N$ exotic solutions have no obvious origins in exceptional
groups.

\bigskip

Our findings are also summarized in Table 1.

\bigskip

Given the well established significance of ${\cal M}_5$ in spacetime
dimension $D=10$ for the two heterotic strings $SO(32)$ and $E_8 \times
E_8$
we are led to observe that for $k=2$ 
(dimensionality ${\cal M}_p$) of $SO(2^p)$ 
in spacetime dimensions $D=2p$ for any of the Mersenne primes,
as
well as the other particular cases listed in our Table 1, the leading
polygonal anomaly ($(p+1)$-agon) is cancelled. With the possibility that
the
non-leading anomalies are also cancelled, we are naturally led to
speculate
that there exist consistent string theories, beyond those presently
established, in the spacetime dimensions and involving the particular
gauge
groups to which we have been led.

\bigskip

This speculation, if verified, will provide one more link between number
theory, particularly the Mersenne primes, and string theory.

\bigskip \bigskip \bigskip

We thank John Schwarz for drawing our attention to ref. \cite{W}. TWK
thanks
PHF and the Department of Physics and Astronomy at UNC Chapel Hill for
their
hospitality while this work was in progress.This work was supported in
part
by the U.S. Department of Energy under Grants No. DE-FG02-97ER41036 and
.DE-FG-5-85ER40226

\newpage

Table 1. Solutions of vanishing leading polygonal gauge anomalies. Given
a $k
$ and $N$, we can find the next value of $k$ (say $k$*) from $(N-k)$ but
the
next N value (say $N$*) corresponding to $k$* requires a calculation. We
have been able to do this calculation up to where a ''?'' appears.
\newline
\underline{Notes:}\newline
$_{N}C_{M}$ is the binomial coefficient
$_{N}C_{M}=N!/(M!(N-M)!)$.\newline
$\dagger $ denotes $D=2p$ where $p$ is a Mersenne prime $M_{p}$ ({\it
c.f.}
Eq.(\ref{primes})).\newline
$\ddagger \ddagger $ denotes the perfect number ${\cal
M}_{p}=2^{p-1}M_{p}$.

\newpage

\begin{center}
Table 1
\end{center}

\bigskip \bigskip

\begin{tabular}{||c|c|c|c||}
\hline\hline
Spacetime dimension(D) & N of SO(N) & $k$ of irrep & dimension of irrep.
\\
\hline\hline
4 & see Ref. \cite{Slansky} & see Ref. \cite{Slansky} & see Ref. \cite
{Slansky} \\ \hline
6$\dagger$ & 8 & 2 & $_8C_2 ={\cal M}_3 \ddagger\ddagger$ \\ \cline{2-4}

& 27 & 6 & $_{27}C_6$ \\ \cline{2-4}
& 98 & 21 & $_{98}C_{21}$ \\ \cline{2-4}
& 363 & 77 & $_{363}C_{77}$ \\ \cline{2-4}
& 1352 & 286 & $_{1352}C_{286}$ \\ \cline{2-4}
& ? & 1064 & ? \\ \hline
8 & 16 & 2 & $_{16}C_2$ \\ \cline{2-4}
& 27 & 3 & $_{27}C_3$ \\ \cline{2-4}
& 147 & 14 & $_{147}C_{14}$ \\ \cline{2-4}
& 256 & 24 & $_{256}C_{14}$ \\ \cline{2-4}
& 1444 & 133 & $_{1444}C_{133}$ \\ \cline{2-4}
& ? & 232 & ? \\ \cline{2-4}
& ? & 1311 & ? \\ \hline
10$\dagger$ & 12 & 4 & $_{12}C_4$ \\ \cline{2-4}
& 32 & 2 & $_{32}C_2 = {\cal M}_5 \ddagger\ddagger$ \\ \cline{2-4}
& 32 & 10 & $_{32}C_{10}$ \\ \hline
12 & N=even & $\frac{N}{2}$ & $_NC_{N/2}$ \\ \cline{2-4}
& $2^6$ & 2 & $_{64}C_2$ \\ \hline
14$\dagger$ & 128 & 2 & $_{128}C_2 = {\cal M}_7 \ddagger\ddagger$ \\
\hline
16 & N=even & $\frac{N}{2}$ & $_NC_{N/2}$ \\ \cline{2-4}
& 27 & 3 & $_{27}C_3$ \\ \cline{2-4}
& $2^8$ & 2 & $_{256}C_2$ \\ \cline{2-4}
& 486 & 3 & $_{486}C_3$ \\ \hline\hline
\end{tabular}

\newpage

\begin{center}
Table 1(continued)
\end{center}

\bigskip \bigskip

\begin{tabular}{||c|c|c|c||}
\hline\hline
Spacetime dimension(D) & N of SO(N) & $k$ of irrep & dimension of irrep.
\\
\hline\hline
18 & $2^9$ & 2 & $_{512}C_2$ \\ \hline
20 & N=even & $\frac{N}{2}$ & $_NC_{N/2}$ \\ \cline{2-4}
& $2^{10}$ & 2 & $_{1024}C_2$ \\ \hline
22$\dagger$ & $2^{11}$ & 2 & $_{2048}C_2={\cal M}_{11}\ddagger\ddagger$
\\
\hline
24 & N=even & $\frac{N}{2}$ & $_NC_{N/2}$ \\ \cline{2-4}
& $2^{12}$ & 2 & $_{4096}C_2$ \\ \hline
26$\dagger$ & $2^{13}$ & 2 & $_{8192}C_2 = {\cal M}_{13}
\ddagger\ddagger$
\\ \hline\hline
D=4n & N=even & $\frac{N}{2}$ & $_{N}C_{N/2}$ \\ \cline{2-4}
& $2^{2n}$ & 2 & $_{4^n}C_2$ \\ \hline
D=4n+2 & $2^{2n+1}$ & 2 & $_{2^{2n+1}}C_2$ \\ \hline
D=2p (p=Mersenne)$\dagger$ & $2^p$ & 2 & ${\cal M}_p \ddagger\ddagger$
\\
34 $\dagger$, 38 $\dagger$, 62 $\dagger$, 178 $\dagger$, 214 $\dagger$,
254 $%
\dagger$ & ? & ? ($k>40$) & ? \\
1042 $\dagger$ & ? & ? ($k>24$) & ? \\
1214 $\dagger$ & ? & ? ($k>15$) & ? \\
2558 $\dagger$ & ? & ? ($k>10$) & ? \\
4406 $\dagger$ & ? & ? ($k>8$) & ? \\
$D=(2p)\dagger\geq4562\dagger$ & ? & ? & ? \\ \hline\hline
\end{tabular}

\end{document}